\documentclass[11pt]{article}
\newcommand{\be}{\begin{equation}}
\newcommand{\ee}{\end{equation}}
\newcommand{\ba}{\begin{eqnarray}}
\newcommand{\ea}{\end{eqnarray}}
\def\L{{\cal L}}
\newcommand{\e}{{\rm e}}
\newcommand{\ep}{\epsilon}
\newcommand{\p}{\partial}
\newcommand{\vev}[1]{\left\langle #1 \right\rangle}
\newcommand{\psibar}{\overline{\psi}}
\newcommand{\dslash}{\hbox{$\partial$\kern-0.5em\raise0.3ex\hbox{/}}}
\def\slash#1{\hbox{$#1$\kern-0.5em\raise0.3ex\hbox{/}}}
\begin{document}
\begin{titlepage}
\rightline{KOBE-TH-01-02}
\rightline{\tt hep-th/0104242}
\vspace{.5cm}
\begin{center}
{\LARGE Off-shell renormalization of \\
\vspace{0.3cm} the massive QED in the unitary gauge}\\
 \renewcommand{\thefootnote}{\fnsymbol{footnote}}
\vspace{1cm} Hidenori SONODA\footnote[2]{E-mail: {\tt
sonoda@phys.sci.kobe-u.ac.jp}}\\
\renewcommand{\thefootnote}{\arabic{footnote}}
\vspace{.2cm}
Physics Department, Kobe University, Kobe 657-8501, Japan\\
\vspace{.2cm} 
April 2001 (revised July 2001)\\
\vspace{.2cm}
\end{center}
\vspace{.3cm}
\begin{abstract}
Despite its simplicity, the unitary gauge is not a popular choice for
practical loop calculations in gauge theories, due to the lack of
off-shell renormalizability.  We study the renormalization properties
of the off-shell Green functions of the elementary electron fields in
the massive QED, in order to elucidate the origin and structure of the
extra ultraviolet divergences which exist only in the unitary gauge.
We find that all these divergences affect the Green functions in a
trivial way such that in coordinate space the off-shell Green
functions are in fact multiplicatively renormalizable.  This result
may generalize to the abelian and non-abelian Higgs theories, for
which the unitary gauge might bring much simplification to the loop
calculations.
\end{abstract}
\end{titlepage}

The unitary gauge is an attractive gauge for the massive gauge
theories for the absence of unphysical degrees of freedom and for the
simplicity of the Feynman rules.  Even though the on-shell S-matrix
elements are known to be calculable in the unitary gauge
\cite{Dublin}, it has not become a popular choice for practical loop
calculations, mainly due to the lack of off-shell
renormalizability.\footnote{The use of the unitary gauge for
calculating the (thermal) effective action was pioneered in
ref.~\cite{DJ}.  The issue of gauge invariance of the effective action
has been resolved in ref.~\cite{TK}.} In the unitary gauge the loop
calculations of the off-shell Green functions give rise to extra
ultraviolet (UV) divergences that cannot be removed by the usual
renormalization of coupling constants and wave functions.  The main
purpose of the present paper is to show that the off-shell Green
functions are in fact renormalizable if we do not consider them in
momentum space, but if we consider them in coordinate space for all
distinct points.  We draw this conclusion from an explicit relation
between the unitary and covariant gauges for the Green functions.  As
a consequence of this relation, we can understand the structure of the
extra divergences.  In this paper we restrict ourselves only to the
simplest case of QED with an explicit photon mass.  Our analysis is
based upon the earlier results on the gauge dependence of QED in the
covariant gauge \cite{Old,S}.

The massive QED is defined by the following gauge invariant
lagrangian\footnote{We will use the euclidean metric throughout.}:
\be \L_{inv} = {1 \over 4} F_{\mu\nu}^2 + {1 \over 2} \left( \p_\mu \phi
+ m A_\mu \right)^2 + \psibar \left( {1 \over i} \dslash - e \slash{A} +
i M \right) \psi \ee
where $\phi$ is a real scalar field called the St\"uckelberg field.  In
the covariant gauge we add a gauge fixing term to obtain the lagrangian
\ba \L_\xi &=& {1 \over 4} F_{\mu\nu}^2 + {1 \over 2} m^2 A_\mu^2 + {1
\over 2 \xi} (\p \cdot A)^2 + {1 \over 2} \left( \p_\mu \phi \right)^2 +
{1 \over 2} \xi m^2 \phi^2 \nonumber\\
&& + \psibar \left( {1 \over i} \dslash - e
\slash{A} + i M \right) \psi \label{Lxi}\ea

We first wish to consider the relation between the covariant gauge and
the unitary gauge.  Using the following gauge invariant variables:
\be B_\mu \equiv A_\mu + {1 \over m} \p_\mu \phi,\quad \psi' \equiv
\e^{i {e \over m} \phi} \psi,\quad \overline{\psi'} \equiv \e^{- i {e
\over m} \phi} \psibar \label{change}\ee
we can rewrite the lagrangian (\ref{Lxi}) as
\be \L_\xi = \L_U (B_\mu, \psi', \overline{\psi'}) + {1 \over 2 \xi m^2}
\left( \left( - \p^2 + \xi m^2\right) \phi - m \p \cdot B \right)^2
\label{Lxi2}\ee
where $\L_U$ is the lagrangian in the unitary gauge:
\be \L_U (B_\mu, \psi, \overline{\psi}) \equiv {1 \over 4} \left( \p_\mu 
B_\nu - \p_\nu B_\mu \right)^2 + {1 \over 2} m^2 B_\mu^2 +
\psibar \left( {1 \over i} \dslash - e \slash{B} + i M \right) \psi \ee
Since $\L_U(B,\psi',\overline{\psi'})$ is independent of $\phi$, we can
first integrate out $\phi$ in the lagrangian (\ref{Lxi2}) if we are only
interested in the Green functions of the gauge invariant fields which
depend only on $B_\mu, \psi', \overline{\psi'}$.  Hence, we
obtain\footnote{Here we use the obvious notation:
$$
\vev{...}_\xi \equiv \int [dAd\psi d\psibar d\phi] ~...~ \e^{- \int_x
\L_\xi (A,\psi,\psibar,\phi)},~~\vev{...}_U \equiv \int [dBd\psi
d\psibar] ~...~ \e^{- \int_x \L_U (B,\psi,\psibar)}
$$
}
\be \vev{\left( A_\mu + {1 \over m} \p_\mu \phi \right) ... \e^{i {e \over
m} \phi} \psi ... \e^{-i {e\over m} \phi} \psibar ...}_\xi
= \vev{B_\mu ... \psi ... \psibar ...}_U \label{xi-ind}\ee
This has the obvious meaning that the Green functions of gauge invariant
fields are independent of $\xi$.  We also note that the contribution of
$\phi$ is a calculable overall factor on the left-hand side, since the
St\"uckelberg field $\phi$ is free.  Hence, Eq.~(\ref{xi-ind}) gives the
explicit $\xi$ dependence of the Green functions in the covariant
gauge \cite{Old,S}.

We have so far dealt with bare fields.  We next wish to consider the
renormalization of the off-shell Green functions in the unitary gauge.
We use the dimensional regularization with $D \equiv 4 - \ep$
dimensional euclidean space.  By renormalizing the fields in the minimal
subtraction (MS) scheme, the lagrangian in the covariant gauge is given
by
\ba \L_\xi &=& {1 \over 4} Z_3 F^2 + {1 \over 2 \xi} \left( \p \cdot A
\right)^2 + {1 \over 2} m^2 A_\mu^2 + {1 \over 2} \left(\p_\mu
\phi\right)^2 + {1 \over 2} \xi m^2 \phi^2 \nonumber\\ &&+ Z_2 \psibar
{1 \over i} \left( \dslash - i e \slash{A} \right) \psi + i Z_2 Z_M M
\psibar \psi \label{Lxiren} \ea
Both $Z_3$ and $Z_M$ are independent of $\xi$, but $Z_2$ depends on
$\xi$ as follows \cite{Gauge}:
\be Z_2 (\xi) = \exp \left[ - \xi {e^2 \over (4 \pi)^2} {2 \over
\ep}\right] Z_2 (0) \label{Z2xidep}\ee
Formally the unitary gauge is obtained by taking the limit $\xi \to
\infty$ in the lagrangian (\ref{Lxiren}), but the limit $\xi \to \infty$
does not commute with minimal subtraction because of the additional
divergences generated in the limit $\xi \to \infty$.  (For example, the
wave function renormalization constant $Z_2(\xi)$ diverges as $\xi \to
\infty$ despite the use of dimensional regularization.)  Hence, we must
first choose the unitary gauge, and then renormalize the fields in the
MS scheme to obtain results free of UV divergences.

By the same change of variables as Eq.~(\ref{change}), we obtain
\be \vev{B_\mu ... \psi ... \psibar ...}_U = \vev{ \left(A_\mu + {1
\over m} \p_\mu \phi \right) ... \e^{i {e \over m} \phi} \psi ... \e^{-
i{e \over m} \phi} \psibar ...}_\xi
\label{relation} \ee
Especially for the electron propagator, we obtain
\ba \vev{\psi(x)\psibar(0)}_U &=&\vev{\e^{i {e \over m} \phi} \psi(x) ~
\e^{- i{e \over m} \phi} \psibar(0)}_\xi \nonumber \\ &=& \e^{- {e^2
\over m^2} \vev{\phi (0) \phi(0)}_\xi} \cdot \e^{{e^2 \over m^2}
\vev{\phi (x) \phi(0)}_\xi } \vev{\psi (x) \psibar(0)}_\xi \label{prop}\ea
Since 
\be \vev{\phi(0) \phi(0)}_\xi = \mu^{\ep \over 2} \int {d^D k \over
(2\pi)^D} {1 \over k^2 + \xi m^2} = {\Gamma \left( -1 + {\ep \over
2}\right) \over (4\pi)^{D \over 2}} \left({\xi m^2 \over
\mu^2}\right)^{- \ep \over 2} \xi m^2, \ee
where $\mu$ is a renormalization mass scale, we must renormalize the
electron propagator further by the factor
\be \exp \left[ - \xi {e^2 \over (4 \pi)^2} {2 \over \ep}\right] \ee
to remove UV divergences entirely.  The $\xi$ dependence (\ref{Z2xidep})
of the wave function renormalization constant implies that in the
unitary gauge the wave function renormalization constant $\tilde{Z}_2$
is the same as the one in the Landau gauge:
\be \tilde{Z}_2 = Z_2 (\xi=0) \label{wave}\ee
As a consequence, the electron field in the unitary gauge has the same
anomalous dimension as the electron field in the Landau gauge.  (It
vanishes at one-loop.)  Thus, in terms of renormalized fields, the
lagrangian for the unitary gauge is given by
\ba \L_U &=& {1 \over 4} Z_3 \left( \p_\mu B_\nu - \p_\nu B_\mu \right)^2
+ {1 \over 2} m^2 B_\mu^2 \nonumber\\
&& + \tilde{Z}_2 \psibar \left( {1 \over i} \dslash - e \slash{B}\right)
\psi +  \tilde{Z}_2 Z_M M i \psibar \psi \ea

{}From Eqs.~(\ref{prop}--\ref{wave}) the renormalized electron
propagator in the unitary gauge and that in the covariant gauge are
related by
\be \vev{\psi (x) \psibar(0)}_U = \e^{- {\xi e^2 \over (4 \pi)^2} \left
( \ln {\xi m^2 \over \bar{\mu}^2} - 1 \right)} \e^{{e^2 \over m^2} \Delta
(x;\xi m^2)} \vev{\psi (x) \psibar (0)}_\xi, \ee
where $\bar{\mu}^2 \equiv 4\pi \mu^2 \e^{- \gamma}$ ($\gamma$ is the
Euler constant), and
\be \Delta (x;\xi m^2) \equiv \int {d^4 k \over (2\pi)^4} {\e^{i k\cdot
x} \over k^2 + \xi m^2} \ee
In general, from Eq.~(\ref{relation}) we obtain
\ba &&\vev{B_\mu ... \psi (y_1) ... \psi (y_N) \psibar (z_1) ... 
\psibar(z_N)}_U \nonumber\\ &=& \e^{- N {\xi e^2 \over (4\pi)^2} \left(
\ln {\xi m^2 \over \bar{\mu}^2} - 1 \right)} \nonumber\\
&&\times \vev{ \left(A_\mu + {1
\over m} \p_\mu \phi\right) ... :\e^{i {e \over m} \phi}: \psi (y_1) ... 
:\e^{- i {e \over m} \phi}: \psibar (z_1) ...}_ \xi \label{Uxi}\ea
where $: \e^{\pm i {e\over m}\phi}:$ denotes normal ordering.  This
gives an explicit relation between the unitary and covariant gauges
for the off-shell Green functions in coordinate space.\footnote{Recall
that the change of the Green functions under an infinitesimal change
of the gauge fixing parameter is given by the Ward identity.  The
above relation (\ref{Uxi}) is the integral of the Ward identity.}

Note that even in the unitary gauge the Green functions in coordinate
space are free of UV divergences by multiplicative renormalization of
the electron fields.  However, the propagator of the free massive
scalar field $\phi$ has the following short-distance singularity
\be \phi (x) \phi (0) = {1 \over 4 \pi^2} {1 \over x^2} + ... \ee
and the exponentiated two-point function $\e^{\pm {e^2 \over m^2}
\vev{\phi (x) \phi(0)}_\xi}$ gives a singularity of order ${1 \over
x^{2 n}}$ at order $e^{2n}$ in perturbation theory.  For $n \ge 2$ the
singularity of order ${1 \over x^{2n}}$ cannot be integrated over
$x=0$, and this unintegrability gives rise to extra UV divergences in
the Fourier transforms of the Green functions.  Despite the elementary
appearance of the electron field in the unitary gauge, it behaves much
as a composite field (of an arbitrarily high scale dimension) with
respect to renormalization.

To clarify the structure of the extra UV divergences in the unitary
gauge, let us study the short-distance singularity between two electron
fields in the covariant gauge.  Up to first order in $e$, we obtain
\be \psi (x) \psibar (0) = {1 \over 4 \pi^2} \left[ 2 i {\slash{x} \over
(x^2)^2} - i M {1 \over x^2} - 2 e {\slash{x} x_\mu \over (x^2)^2} A_\mu
+ ... \right] \ee
This implies that $\e^{- i p\cdot x}:\e^{i {e \over m} \phi (x)}: \psi
(x)~\times~:\e^{- i{e \over m} \phi(0)}: \psibar (0)$ contains the
following unintegrable singularity up to order $e^3$ in perturbation
theory:
\be {1 \over (4\pi^2)^2} {1 \over x^4} {e^2 \over m^2} \left( {1 \over
2} \slash{p} - i M - {e \over 2} \slash{A} \right) \ee
Thus, we expect that the following UV singularities result even after
multiplicative renormalization of fields if we consider the Fourier
transforms of the Green functions:
\ba && \int d^D x~\e^{- i p \cdot x} \vev{\psi (x) \psibar (0)}_U
\nonumber\\ &=& {\rm UV~finite} + {1 \over (4\pi)^2} {2 \over \ep}~{e^2
\over m^2} \left( {1 \over 2} \slash{p} - i M \right)\\ && \int d^D
x~\e^{- i p \cdot x - i k \cdot y} \vev{\psi (x) \psibar (0) A_\mu
(y)}_U \nonumber\\ &=& {\rm UV~finite} + {1 \over (4\pi)^2} {2 \over
\ep}~{e^2 \over m^2} {- e \over 2} \int d^D y~\e^{- i k \cdot y}
\vev{\slash{A} (0) A_\mu (y)}_U \ea
These results have been checked explicitly by one-loop calculations in
the unitary gauge.  Analogously, at order $e^4$ the short-distance
singularity
\be :\e^{i {e\over m}\phi (x)}: \psi (x)~\times~ :\e^{i {e\over m}\phi
(0)}: \psi (0) \sim - {1 \over 2} {e^4 \over m^4} {1 \over (4 \pi^2)^2}
{1 \over x^4} :\e^{ 2 i{e \over m} \phi}: \psi \psi (0) \ee
gives rise to a pole at $\ep = 0$ in the Fourier transforms of Green
functions involving two $\psi$ fields.  The product of three or more
electron fields also generates unintegrable singularities.  The
structure of the extra singularities is the same as in the Green
functions of composite fields in the covariant gauge.

Now that we understand the origin and structure of the extra UV
divergences in the unitary gauge, we can easily conclude that they are
harmless.  Clearly the extra divergences do not affect the S-matrix,
since the wave packets for the asymptotic particle states have no
overlap in space, and the extra divergences that occur only if two or
more fields coincide in space are irrelevant.  Thus, the UV finiteness
of the Green functions for distinct space coordinates guarantee the UV
finiteness of the S-matrix.

Not everything we have found for the massive QED in the unitary gauge
generalizes to the abelian and non-abelian Higgs theories in the unitary
gauge.  The free St\"uckelberg field $\phi$ is replaced by an
interacting phase of the Higgs field in the latter theories, and the
simple formula such as (\ref{Uxi}) has no counterpart.

What generalizes is the qualitative feature of the renormalization
properties.  We expect that the unitary gauge is renormalizable in
coordinate space, and the extra UV divergences of the Green functions in
the momentum space are due to two or more fields coincident in space.

The massive QED that we have studied in this paper is a very simple
theory treated in the covariant gauge, thanks to the decoupling of the
St\"uckelberg field $\phi$.  The unitary gauge does not bring any
simplification either to the number of degrees of freedom or to the
Feynman rules.  On the contrary, vast simplification is expected of the
unitary gauge for both the abelian and non-abelian Higgs theories, for
which the scalar and Faddeev-Popov ghosts are highly interacting in the
covariant (or $R_\xi$) gauge.

The extra UV divergences in the unitary gauge are absent in the S-matrix
elements \cite{Dublin}, and in principle one can perform perturbative
loop calculations without worrying about the divergences that appear
before taking the mass-shell limit.  For this procedure to gain popular
acceptance, however, good understanding of the origin and structure of
UV divergences is necessary for both abelian and non-abelian Higgs
theories.

\vspace{0.4cm}
\noindent{\Large \bf Acknowledgment}

This work was supported in part by the Grant-In-Aid for Scientific
Research from the Ministry of Education, Science, and Culture, Japan
(\#11640279).

\end{document}